\documentstyle[prl,aps]{revtex}

\begin{document}

\draft

\preprint{}

\def\be{\begin{equation}}

\def\ee{\end{equation}}

\def\bea{\begin{eqnarray}}

\def\eea{\end{eqnarray}}

\title{ Quasi-long-range order in nematics confined in random
porous media}

\author{D.E.  Feldman}

\address{ Condensed Matter Physics Department, Weizmann Institute of Science,
76100 Rehovot, Israel\\ Landau Institute for Theoretical Physics, 142432
Chernogolovka, Moscow region, Russia }

\maketitle

\begin{abstract}

We study the effect of random porous matrices on the ordering in nematic liquid
crystals.  The randomness destroys orientational long-range order and drives the
liquid crystal into a glass state.  We predict two glass phases one of which
possesses quasi-long-range order.  In this state the correlation length is
infinite and the correlation function of the order parameter obeys a power
dependence on the distance.  The small-angle light-scattering amplitude diverges
but slower than in the bulk nematic.  In the uniaxially strained porous matrices
two new phases emerge.  One type of strain induces an anisotropic
quasi-long-range-ordered state while the other stabilizes nematic long-range
order.

\end{abstract}


Quenched disorder is inevitably present even in the most pure solids.  This
explains a lot of phenomena, e.g.  the residual resistance of metals.  On the
other hand, liquids are usually homogeneous and introducing quenched disorder in
them requires special efforts.  One of the approaches consists in pouring a
liquid into a randomly interconnected network of pores.  Such
liquid-porous-matrix systems emerge in many natural and technological processes
giving rise to a lasting scientific activity.  The recent surge of interest in
the field is due to a new micropore material:  silica aerogel.  Its density can
be varied in a wide range up to more than 99\% void volume fraction.  This
allows the investigation of both strongly and weakly confined fluids.  The most
interesting situation emerges in systems with many degrees of freedom, e.g.
He-3 \cite{He3} and liquid crystals \cite{Wu92,ammy,b92,ZKCWG,WZG,DNMR}.  In
these substances the porous matrix not only geometrically confines the liquid
but also induces a random orienting field that fixes the direction of the order
parameter near the surface of the matrix.  The random-field (RF) disorder is
known to cause spin-glass effects \cite{RN} and such phenomena were indeed
observed experimentally in liquid-crystal-aerogel systems.  In particular, a
slow glassy dynamics was reported in Refs.  \cite{Wu92,ammy}.  Another effect of
the disorder is the suppression of the isotropic-nematic phase transition.

In many experiments \cite{Wu92,b92,ZKCWG} the sharp isotropic-nematic transition
inherent to the bulk liquid crystal was substituted by continuous ordering.  The
genuine phase transition was observed \cite{WZG} only in highly porous aerogels.
However, even these systems do not have nematic long-range order (LRO) as
follows from the Imry-Ma argument \cite{IM,Kats} and the results of deuteron NMR
measurements \cite{DNMR}.  Hence, a question arises about the nature of the new
phase emerging below the transition.  This issue was addressed in several recent
numerical simulations \cite{JCP,PRL95,PRL94,CH}.  Most of them used simplified
models with a discrete symmetry \cite{JCP,PRL95,PRL94}.  Although the structure
of the phase diagram \cite{JCP,PRL95,PRL94} agrees with the experiments the
nematic LRO emerging in these models is an artifact of the discrete symmetry
which allows LRO in three-dimensional RF systems \cite{IM}.  An attempt to
investigate numerically a more realistic continuous model \cite{CH} suggested an
interesting scenario of quasi-long-range order (QLRO) that is the infinite
correlation length and power dependences of the correlation functions on the
distance.  However, the results of Ref.  \cite{CH} should be taken with care
since the Imry-Ma domain length \cite{Saslow} is comparable with the sample size
used for this simulation.

In the absence of reliable numerical results analytical approaches are
especially desirable.  However, the existing theory \cite{PRL94} does not extend
beyond the mean field approximation.  As in the other RF problems it
underestimates fluctuations and incorrectly predicts nematic LRO.  The present
paper for the first time studies the low-temperature phase of disordered
nematics beyond the mean field theory.  We apply the functional renormalization
group (RG) in $4-\epsilon$ dimensions.  RG flows have different character
depending on the type of ordering.  LRO corresponds to the fixed point in which
the disorder and temperature are zero.  If there is only short-range order the
RG flow enters a strong coupling regime.  A non-trivial fixed point describes
QLRO.  It was known for a long time that QLRO is possible in two-dimensional
pure systems with the Abelian symmetry \cite{2DXY}.  Recently the same type of
ordering was discovered in three-dimensional RF systems \cite{RFXY1,RFXY2}.  In
these systems QLRO is not prohibited by the non-Abelian symmetry \cite{PRB}.
Thus, QLRO is more common in disordered systems than clean ones.  We demonstrate
that QLRO emerges in our problem whereas LRO is absent.  An experimental
signature of this ordering is the divergence of the small-angle light-scattering
cross-section (\ref{10}).

The free energy density $F=F_{\rm d}+F_{\rm pm}$ of the nematic in the porous
matrix includes the Frank distortion energy \cite{dG} $F_{\rm d}=[K_1({\rm
div}{\bf n})^2+K_2({\bf n}{\rm curl}{\bf n})^2+K_3({\bf n}\times{\rm curl}{\bf
n})^2]/2$, where ${\bf n}$ is the director, and the interaction $F_{\rm pm}$
with the surface of the random matrix.  The interaction tends to align the
director parallel to the surface \cite{star}.  We model the interaction as
$F_{\rm pm}=({\bf hn})^2$, where ${\bf h}$ is a random vector representing the
normal to the surface.  This is the simplest choice compatible with the
equivalence of the opposite orientations of the director.  Due to the
universality the model captures all large-scale physics and allows quantitative
prediction of the critical exponents in the QLRO state.  The average amplitude
of the random vector ${\bf h}$ is a measure of the disorder strength.  It is a
phenomenological parameter which depends on the pore size, anchoring energy and
fractal structure of the porous matrix.  The microscopic expression for ${\bf
h}$ is discussed in Ref.  \cite{RT}.

In the one-constant approximation $K_1=K_2=K_3$ the energy $F=F_{\rm d}+F_{\rm
pm}$ reduces to the Hamiltonian of the random-anisotropy (RA) Heisenberg model
\cite{PRB}.  Since that model possesses QLRO the same ordering is expected for
the randomly confined nematic.  However, in all nematics $K_1, K_3>K_2$ and this
could change the critical exponents of the correlation functions in the QLRO
state in comparison with the random Heisenberg model.  Below we demonstrate that
this is not the case, i.e.  the nematic in the porous matrix belongs to the
universality class of the RA Heisenberg model.  To get a simple idea why it
occurs we first consider a two-dimensional nematic film with the director ${\bf
n}=(n_{\rm x}, n_{\rm y}, n_{\rm z})=(\cos\phi , \sin\phi , 0)$ confined in the
plane $xy$ of the film in the absence of the disorder.  The Frank energy is
$F_d=(K_1+K_3)({\bf \nabla}\phi)^2/2 +(K_3-K_1)\{\cos
2\phi[(\partial_x\phi)^2-(\partial_y\phi)^2]/2 +\sin
2\phi\partial_x\phi\partial_y\phi \}$.  The low-temperature phase of this system
possesses QLRO, only the term $(K_1+K_3)({\bf \nabla}\phi)^2/2$ being relevant
at the large scales since $\langle\sin 2\phi\rangle=\langle\cos 2\phi\rangle=0$

The systematic consideration is based on the RG equations in $4-\epsilon$
dimensions.  Our method follows the line of Ref.  \cite{PRB} and is briefly
described bellow.  It is convenient to eliminate the disorder with the replica
trick \cite{MPV}.  We search for a zero-temperature fixed point.  We ascribe the
scaling dimension 0 to the director ${\bf n}$.  Then the scaling dimension of
the temperature $\Delta_T=-2+O(\epsilon)$.  Any term of the effective replica
Hamiltonian containing $m$ different replica indices is proportional \cite{DF}
to $1/T^{m-1}$.  This allows us to show that all operators with three or more
different replica indices are irrelevant \cite{DF}.  Hence, all relevant
operators of the appropriate symmetry are included in the following replica
Hamiltonian

\begin{equation} \label{1} H_R=\int d^3 r \left[
\frac{1}{2}\sum_a(\lambda_1\partial_\alpha n^a_\beta\partial_\alpha n^a_\beta+
\lambda_2\partial_\alpha n^a_\alpha\partial_\beta n^a_\beta + \lambda_3
n^a_\alpha\partial_\alpha n^a_\beta n^a_\gamma\partial_\gamma n^a_\beta)
-\sum_{ab}\frac{R({\bf n}^a{\bf n}^b)}{T} \right], \end{equation} where $a,b$
are replica indices, $\alpha,\beta=x,y,z$ label the spatial coordinates,
$\lambda_1=K_2, \lambda_2=K_1-K_2,\lambda_3=K_3-K_2$, $T$ is the temperature,
the function $R(z)$ describes the disorder, and the summation over the repeated
indices $\alpha$ and $\beta$ is assumed.  Due to the symmetry ${\bf
n}^a\leftrightarrow -{\bf n}^a$ the function $R(z)$ is even.  Below we measure
the temperature in units of $K_2$, and hence set $\lambda_1=1$.  To define the
energy in $4-\epsilon$ dimensions we add to the Hamiltonian (\ref{1}) the term
$\lambda_0\sum_{\alpha\beta}\partial_\alpha n^a_\beta \partial_\alpha
n^a_\beta/2$, where $\alpha$ labels the coordinates in the
$(1-\epsilon)$-dimensional subspace, $\beta=x,y,z$.  The stability conditions
\cite{dG} $K_1, K_3>0$ lead to the inequality

\be \label{stability} \lambda_2,\lambda_3>-1.  \ee

We represent each replica ${\bf n}^a({\bf x})$ of the director as a combination
of small-scale fields $\phi_i^a({\bf x}), i=1,2$ and a large-scale field ${\bf
n}'^a({\bf x})$ of the unit length:

\begin{equation} \label{2} {\bf n}^a({\bf x})={\bf n}'^a({\bf
x})\sqrt{1-\sum_i(\phi_i^a({\bf x}))^2}+ \sum_i\phi_i^a({\bf x}){\bf e}_i^a({\bf
x}), \end{equation} where the unit vectors ${\bf e}_i^a({\bf x})$ are
perpendicular to each other and the vector ${\bf n}'^a({\bf x})$.  The fields
$\phi_i$ change at small scales $a<r<L$, where $a$ is the molecule size, $L\gg
a$.  The field ${\bf n}'$ changes at the scales $r>L$.  The RG procedure
consists in integrating out the small-scale fields $\phi_i$ and the rescaling
such that the effective Hamiltonian of the field ${\bf n}'$ would have the
structure (\ref{1}) with new constants.  The rescaling is defined in such a way
that $\lambda_1=1$ remains unchanged.  The RG equations in the first order in
$\epsilon=4-D$ read

\begin{equation} \label{4} \frac{dT}{d\ln L}=-(D-2)T + (1-\lambda_3)C_\phi T;
\frac{d\lambda_2}{d\ln L}=-\lambda_2(1+\lambda_3)C_\phi; \frac{d\lambda_3}{d\ln
L}= -(3\lambda_3+\lambda_3^2 - \lambda_2)C_\phi, \end{equation} where the
constant

\begin{equation} \label{6}
C_\phi=\frac{dR(z=1)/dz}{8\pi^2\sqrt{\lambda_0(1+\lambda_3)}}\left[1+\frac{1}{1+\lambda_2}\right]
\end{equation} describes the fluctuations of the small-scale fields

\begin{equation} \label{7}
\langle\phi_1^2\rangle=\langle\phi_2^2\rangle=C_\phi\ln(L/a).  \end{equation} We
omit the RG equations for $\lambda_0$ and $R(z)$ since their structure is
irrelevant below.  Eqs.  (\ref{4}) have the only fixed point compatible with the
stability conditions (\ref{stability}).  In this fixed point
$T=\lambda_2=\lambda_3=0$ and Eq.  (\ref{1}) reduces to the Hamiltonian of the
RA Heisenberg model which thus describes the large-distance physics of the
randomly confined nematic.  Since that model possesses QLRO in its
low-temperature phase for weak disorder, QLRO is also possible in confined
nematics.  For strong disorder or high temperature the ordering disappears.
Thus, there are three phases:  the high-temperature isotropic phase and two
low-temperature glass phases with and without QLRO.  In both glass phases the
local orientation of the director is fixed by the random potential.  As
discussed below the disorder driven transition between the glass phases is
related with topological defects.

The large-scale correlations of the director lead to strong small-angle light
scattering.  We determine its intensity in the limit of the weak optical
anisotropy, i.e.  assuming that in the dielectric tensor
$\epsilon_{\alpha\beta}=\epsilon_\perp\delta_{\alpha\beta}+\epsilon_an_\alpha
n_\beta$ the anisotropic term $\epsilon_a\ll\epsilon_\perp$.  In this case the
scattering cross-section can be found with the Born approximation.  The
scattering cross-section with the change of the wave vector by ${\bf q}$ is
given by the expression \cite{dG}

\begin{equation} \label{8} \sigma({\bf q})=|\omega^2/(4\pi
c^2)i_\alpha\epsilon_{\alpha\beta}({\bf q})f_b|^2, \end{equation} where $\omega$
is the light frequency, ${\bf i}$ and ${\bf f}$ are the unit vectors specifying
the initial and final polarizations, $\epsilon_{\alpha\beta}({\bf q})$ is the
Fourier transform of the dielectric tensor.  Hence, $\sigma({\bf q})\sim\langle
Q_{\alpha\beta}({\bf q})Q_{\alpha\beta}(-{\bf q})\rangle$, where
$Q_{\alpha\beta}=n_\alpha n_\beta-\delta_{\alpha\beta}/3$ is the order parameter
and the angular brackets denote the disorder and thermal averages.  In contrast
to the bulk nematic the scattering is caused not by the thermal fluctuations but
by the frozen configuration of the director.  The cross-section $\sigma({\bf
q})$ is proportional to the Fourier transform of the correlation function
$G({\bf r})=\langle Q_{\alpha\beta}({\bf 0})Q_{\alpha\beta}({\bf r})\rangle$.
In the QLRO state this correlator obeys a power dependence on the distance
$G(r)\sim r^{-\eta}$.  To calculate the exponent $\eta$ we decompose
$Q_{\alpha\beta}$ into small-scale and large-scale parts with Eq.  (\ref{2})
and average over the small-scale fluctuations with Eq.  (\ref{7}):

\begin{eqnarray} \label{9} \langle Q_{\alpha\beta}({\bf 0})Q_{\alpha\beta}({\bf
r})\rangle_\phi= \{n'_\alpha({\bf 0})n'_\beta({\bf
0})(1-\sum_i\langle\phi^2_i\rangle)+\sum_{ij} e_\alpha^i({\bf 0})e^j_\beta({\bf
0})\langle\phi_i\phi_j\rangle- \delta_{\alpha\beta}/{3}\}\times & & \nonumber \\
\{n'_\alpha({\bf r})n'_\beta({\bf r})(1-\sum_i\langle\phi^2_i\rangle)+\sum_{ij}
e_\alpha^i({\bf r})e^j_\beta({\bf r})\langle\phi_i\phi_j\rangle
-\delta_{\alpha\beta}/{3}\} = Q'_{\alpha\beta}({\bf 0})Q'_{\alpha\beta}({\bf
r})[1-6C_\phi\ln L/a], & & \end{eqnarray} where $Q'_{\alpha\beta}=n'_\alpha
n'_\beta-\delta_{\alpha\beta}/3$, $\langle ...  \rangle$ denotes the average
over the fluctuations of $\phi$, and the relation
$\langle\phi_i\phi_j\rangle\sim\delta_{ij}$ which is valid in the RA Heisenberg
fixed point is used.  The constant $C_\phi=0.309\epsilon$ Eq.  (\ref{7}) is the
same as in the fixed point of the RA Heisenberg model \cite{PRB}.  The exponent
$\eta$ can be found with the iterative use of Eq.  (\ref{9}) at each RG step
until the scale $L=r$ is reached.  At the scale $r$ the values of the
renormalized director field ${\bf n}'$ are the same at the points ${\bf 0}$ and
${\bf r}$.  Hence, $Q'_{\alpha\beta}({\bf 0})Q'_{\alpha\beta}({\bf r})\sim 1$
and $r^{-\eta}\sim (1-6C_\phi\ln L/a)^K$, where $K=\ln (r/a)/\ln (L/a)$ is the
number of the RG steps.  Thus, $\eta=6C_\phi$.  The small-angle scattering
cross-section is given by the expression

\begin{equation} \label{10} \sigma({\bf q})\sim q^{-D+\eta}=q^{-4+2.9\epsilon}.
\end{equation}

The uniaxial stress modifies the large-distance behavior.  The compression along
the $z$-axis can be described by adding to the Hamiltonian the term
$F_S=An_z^2$, where $A>0$, since the deformation tends to make the pore surfaces
parallel to the $xy$ plane and hence favors the planar configuration of the
director.  The uniaxial stretch is described by $F_S=An_z^2$ with a negative
$A$.  In both cases $A$ is proportional to the deformation.  The effect of the
electric field is analogous to the effect of the stress but the sign of the
electric energy \cite{dG} $F_e=-\epsilon_a({\bf nE})^2/8\pi$ is fixed for a
given substance.  The RG flow is unstable with respect to the perturbation $F_s$
and new regimes emerge at the scale $R=R_c$ at which the renormalized $A(R)\sim
1$.  The critical length $R_c$ can be found analogously to the correlation
length of the RA Heisenberg model in the uniform magnetic field \cite{PRB}.  At
small $A$ the result is $R_c\sim |A|^{-1/(2-2C_\phi)}=|A|^{-0.5-0.15\epsilon}$.
The stretched system is long-range-ordered at the scales $R>R_c$.  The nematic
order parameter can be calculated analogously to the magnetization of the RA
Heisenberg model in the uniform magnetic field and is given by the formula
$Q=\langle n_\alpha n_\beta-\delta_{\alpha\beta}/3\rangle\sim
R_c^{-3C_\phi}\sim|A|^{0.46\epsilon}$.  LRO can also be achieved by applying an
arbitrarily weak external magnetic field to the confined nematic since the
magnetic contribution to the energy \cite{dG} $F_m=-\chi_a({\bf nH})^2/2$ has
the same structure as the energy related with the uniaxial stretch.  A more
interesting situation emerges under the compression.  The director averaged over
a scale $R>R_c$ is confined in the $xy$-plane.  The system is thus described by
the RA XY model.  It possesses QLRO but the critical exponents are different
from the exponents of the Heisenberg model.  Thus, at the scale $R_c$ the
cross-over from one QLRO state to another occurs.  Using the RA XY fixed point
found in Ref.  \cite{PRB} and repeating the derivation of Eq.  (\ref{10}) one
finds the Born light-scattering cross-section for $q<1/R_c$:  $\sigma({\bf
q})\sim q^{-4+\epsilon(1+\pi^2/9)}.$ In the RA XY regime the cross-section
(\ref{8}) is anisotropic:  the small-angle scattering is suppressed, if the
incident or scattered light is polarized along the compression direction.

Our RG procedure is based on the decomposition (\ref{2}) which makes sense only
if the director change is slow at the microscopic scale $a$.  This condition is
broken in the core of a topological defect.  However, the topological defects
are irrelevant at small $\epsilon=4-D$ for weak disorder.  This can be
understood from the consideration of the contribution of the disclination loops
and the pairs of point defects of size $l\gg a$ to the RG equations at the scale
$l$.  After averaging over the small-scale fluctuations the size $l$ of the
topological excitations plays the role of the ultra-violet cut-off.  The
renormalized temperature is small:  $T(l)\ll 1$.  Hence, the thermal
fluctuations are irrelevant.  The disorder-induced term $R({\bf n}_a{\bf
n}_b)\sim\epsilon$ in the renormalized replica Hamiltonian (\ref{1}) is of order
$\langle h^4\rangle$, where the random vector ${\bf h}$ describes the
(renormalized) random anisotropy $F_{pm}=({\bf hn})^2$, $\langle ...  \rangle$
denotes the average over the realizations of the disorder.  The excitation
energy can be compensated by the interaction with the disorder only in the
positions where $h\sim 1$.  The concentration of such positions is exponentially
small $\sim\exp(-1/\epsilon)$.  Hence, the defects produce the corrections of
order $\exp(-1/\epsilon)$ to the RG equations and do not modify the results of
the paper qualitatively.  The concentration of the topological excitations of
size $l$ is of order $l^{-D}\exp(-1/\epsilon)$.  The above discussion is valid,
if the disorder is weak.  In the case of the strong disorder the topological
defects are present at the microscopic scale $a$ and QLRO is absent.  Thus, the
topological defects drive the system into another glass state in which the
orientation of the director is determined only by the local random potential.

Recently it was suggested that the RF model (\ref{1}) describes nematic
elastomers \cite{FT}.  However, it is unclear if this model and hence the above
results are applicable to nematic elastomers because of elasticity-mediated
non-local interactions in these substances \cite{E}.  The non-local dipole
interactions are present also in the amorphous magnets which could be described
by the RA Heisenberg model in the absence of the dipole forces.  The effect of
the long-range interactions on the stability of QLRO is an open question.  Such
forces are absent in randomly confined nematics which provide a genuine
realization of the RA Heisenberg model and can be used for an experimental test
of the possibility of QLRO in the non-Abelian systems.  In conclusion, we have
demonstrated that weakly disordered nematics possess QLRO in their
low-temperature phase.  The uniaxial stretch and the external magnetic field
stabilize LRO while the uniaxial compression drives the liquid crystal into
another QLRO state.

The author thanks A.  Kamenev, V.V.  Lebedev, V.P.  Mineev and V.  Steinberg for
useful discussions and R.  Whitney for the critical reading of the manuscript.


\begin{references}
\bibitem{star} Other types of anchoring do not modify the results of the paper.
\end{references}
 \end{document}